# XPS evidence of degradation mechanism in hybrid halide perovskites

I.S. Zhidkov[1], A.I. Poteryaev[2], A. Kukharenko[1], L.D. Finkelstein[2], S.O. Cholakh,
A.F. Akbulatov[3], P.A. Troshin[3,4], C.-C. Chueh[5] and E.Z. Kurmaev[1,2]

[1]*Institute of Physics and Technology, Ural Federal University, Mira 9 str., 620002 Yekaterinburg, Russia*
[2]*M.N.Mikheev Institute of Metal Physics of Ural Branch of Russian Academy of Sciences, S.Kovalevskoi 18 str., 620108 Yekaterinburg, Russia*
[3]*Institute for Problems of Chemical Physics of the Russian Academy of Sciences (ICP RAS), Semenov prospect 1, Chernogolovka, 142432, Russia*
[4]*Skolkovo Institute of Science and Technology, Nobel street 3, Moscow, 143026, Russia*
[5]*Advanced Research Center for Green Materials Science and Technology, National Taiwan University Taipei 10617, Taiwan*

Abstract. The paper presents the results of measurements of XPS valence band spectra of $SiO_2/MAPbI_3$ hybrid perovskites subjected to irradiation with visible light and annealing at an exposure of 0-1000 hours. It is found from XPS survey spectra that in both cases (irradiation and annealing) a decrease in the I:Pb ratio is observed with aging time, which unambiguously indicates $PbI_2$ phase separation as a photo and thermal product of degradation. The comparison of the XPS valence band spectra of irradiated and annealed perovskites with density functional theory calculations of the $MAPbI_3$ and $PbI_2$ compounds have showed a systematic decrease in the contribution of I 5p-states and allowed us to determine the threshold for degradation, which is 500 hours for light irradiation and 200 hours for annealing.

1. Introduction

Hybrid halide perovskites $MAPbX_3$ ($MA=CH_3NH_3$, X=Cl, Br, I) made a real revolution in photovoltaics and have now achieved outstanding efficiency comparable to and even greater than that in silicon [1]. The main problem now is to overcome their limited service life under operating conditions and improve their photo and thermal stability [2-4]. This is due to the fact that a fundamental understanding of the physicochemical processes that cause degradation is necessary for the development of perovskite solar cells with a high resource of their practical use. It is currently well established that external factors such as light, temperature, humidity and $O_2$ can contribute to degradation processes [5-7]. One way of protecting against exposure to humidity and oxygen of ambient atmosphere is encapsulation of hybrid perovskites, which has already shown its effectiveness [8]. However, the problem of degradation caused by exposure to visible light and temperature, that are factors present in actual use of photovoltaic devices, still remains. To investigate the mechanisms of degradation in hybrid perovskites, the various methods were used including x-ray diffraction [9-10], electron microscopy [11], thermo gravimetry analysis [9], photoluminescence [12], and first-principle calculations [13] of the formation energies of various structural defects. In this article, the mechanism of photo and thermal degradation of methylammonium lead iodide - $MAPbI_3$ is investigated using x-ray photoelectron valence band spectra which are found to be an effective tool to probe the chemical bonding and electronic structure of solids and their comparison with specially performed density functional theory (DFT) calculations.

## 2. Experimental and Calculation Details

Glass substrates (5 Ω, Luminescence Technology Corp.) were sequentially cleaned with toluene and acetone and sonicated in deionized water, acetone and isopropanol. The $MAPbX_3$ precursor solutions in DMF (~ 45 w. %) were spin-coated at 5000 rpm inside a nitrogen glove box. The toluene (200 μL) was dropped on the film 4-5 s after initiation of the spin-coating, inducing the film crystallization. Spinning was continued for 45 s and then the deposited films were annealed at 100˚C for 15 min on a hotplate installed inside the glove box. The thermal aging experiments were conducted inside a nitrogen MBraun glovebox with $O^2$<0.1 ppm and $H_2O$<0.1 ppm using a calibrated hot plate as a heat source. The samples were placed on the hotplate at 90 ºC and covered with a non-transparent lid to avoid exposure to the ambient light. The photochemical aging experiments were performed using specially designed setups integrated with the dedicated MBraun glove box using LG sulfur plasma lamp as a standard light source, which is known to provide a good approximation of the solar AM1.5G spectrum. The light power at the samples was ~70 mW/cm$^2$, while the temperature was 45±2 ºC (provided by intense fan cooling of the sample stage).

X-ray photoelectron spectroscopy (XPS) was used to measure valence band spectra (VB) with help of a PHI XPS 5000 Versaprobe-spectrometer (ULVAC-Physical Electronics, USA) with a spherical quartz monochromator and an energy analyzer working in the range of binding energies (BE) from 0 to 1500 eV. The energy resolution was ∆E ≤ 0.5 eV. The samples were kept in the vacuum chamber for 24 h prior to the experiments and were measured at a pressure of $10^{-7}$ Pa.

Electronic structure calculations were performed using VASP code[14] with the standard frozen-core projector augmented-wave (PAW) method. The cut-off energy for basis functions was set to 500 eV. Perdew, Burke, Ernzerhof's generalized gradient approximation was used for exchange-correlation [15]. Atomic positions are relaxed since all the forces on atoms are below 0.05eV/A˚. The spin-orbital coupling was switched on due to strong relativistic effect of Pb. 76 and 196 k-points were utilized in the irreducible part of the first Brilluoin zone for $CH_3NH_3PbI_3$ and $PbI_2$, respectively.

### 3. Results and Discussion

As shown by X-ray diffraction (XRD) measurements [10], the decomposition of $MAPbI_3$ revealed the disappearance of diffraction peaks, which are characteristics of $MAPbI_3$, and the appearance of $PbI_2$ peaks, which suggests the following reaction:

$$CH_3NH_3PbI_3 \rightarrow PbI_{2\,solid} + CH_3NH_3I \rightarrow PbI_{2\,solid} + CH_3I_{gas} + NH_{3\,gas}$$

To study the mechanism of photo and thermal degradation of perovskite, we first measured the XPS survey spectra of irradiated and annealed $MAPbI_3$ samples with aging time of 0-1000 hours in the energy range of 0-600 eV. These data are presented in Fig. 1, and the surface compositions determined from these spectra (a quantitative estimate of the relative concentration of elemental species is obtained from the measured spectroscopic intensities, corrected by the elemental and orbital specific photoemission cross sections) are shown in the Table (in at.%).

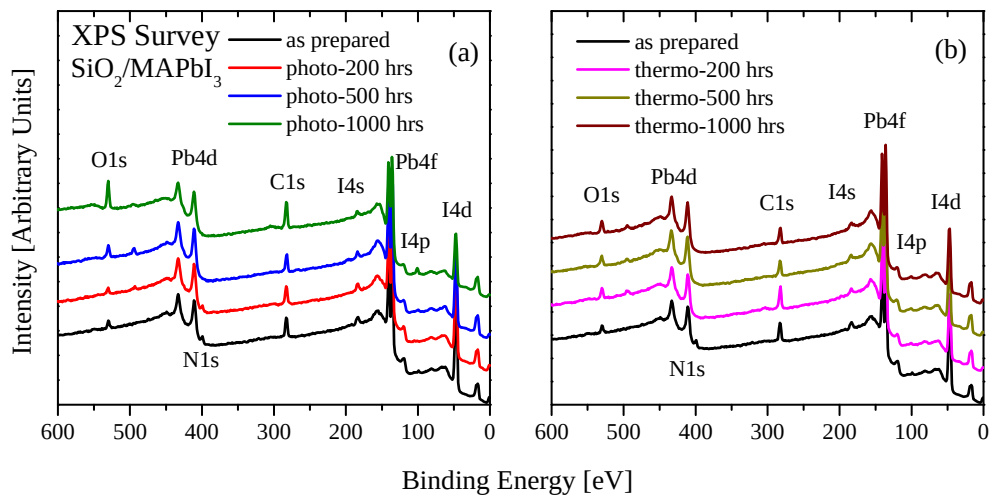

Fig. 1. XPS survey spectra of as-prepared, light irradiated (a) and annealed (b) $SiO_2/MAPbI_3$.

Table. Surface composition of as-prepared, light irradiated and annealed $SiO_2/MAPbI_3$ (in at.%)

| $SiO_2/MAPbI_3$ | C | O | I | Pb | N | Si | I:Pb |
|---|---|---|---|---|---|---|---|
| As prepared | 44,9 | 6,7 | 27,5 | 10,3 | 8,3 | 2,1 | 2,66 |
| Photo 200 h | 39,5 | 6,1 | 32,6 | 13,7 | 5,5 | - | 2.37 |
| Photo 500 h | 37 | 15,1 | 25,6 | 14,8 | - | 3,2 | 1,72 |
| Photo 1000 h | 50,8 | 20,2 | 11,5 | 8,9 | - | 6,6 | 1.29 |
| Thermo 200 h | 51 | 9,4 | 21,4 | 12,9 | - | 2,4 | 1,65 |
| Thermo 500 h | 37,7 | 10,4 | 27,3 | 16,3 | - | 4 | 1,67 |
| Thermo 1000 h | 40 | 13,3 | 24 | 14,3 | - | 4,5 | 1,67 |

As it follows from this data, only elements belonging to perovskite (C, N, Pb and I) and the substrate (Si and O) were found on the surface of the studied samples. To test the proposed mechanism for the degradation of hybrid halide perovskite involving the precipitation of a $PbI_2$ phase, it is important to determine the I:Pb ratio and its evolution with the time of irradiation and annealing. These ratios presented in the Table show their consistent decrease when increasing the dose of radiation with visible light and annealing from 0 to 1000 hours. However, if the general trend of a change of I:Pb ratio in attitude is preserved over aging time for both external influences, then the threshold values of degradation differ significantly. So for irradiation with visible light, the threshold degradation begins with a time of 500 hours (I:Pb=1.72) whereas for annealing already at 200 hours (I:Pb=1.65).

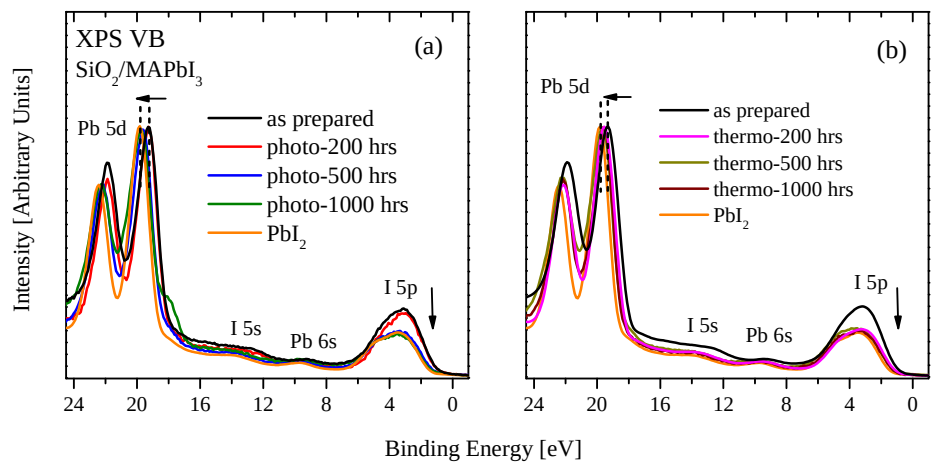

Fig. 2. XPS Pb 5d+VB spectra of as-prepared, light irradiated (a) and annealed (b) $SiO_2/MAPbI_3$.

Now let's see how the change in the I:Pb ratio affects the XPS valence band spectra. Since the degradation mechanism described above assumes the unchanged lead content in perovskite, the data shown in Figure 2 (a-b) were normalized to the intensity of the XPS Pb 5d-spectra. As follows from these spectra, an increase in the exposure time actually leads to a decrease in the relative intensity of the main band located at 0-7 eV reflecting the distribution of I 5p-states, which actually means a decrease in the contribution of these states with aging time. Here, the reference is the XPS VB spectrum of the $PbI_2$ compound which is the final product of the photo and thermal decomposition of the $MAPbI_3$ perovskite, and as a result we see that at the limiting aging time (1000 hours), the XPS VB spectra of the irradiated and annealed samples actually coincide with those of the $PbI_2$ compound. It is curious that the threshold values of degradation expressed in reducing the contribution of I 5p-states exactly coincide with those determined from the I:Pb ratios (see Table) and correspond to 500 hours for irradiation with visible light and 200 hours for annealing.

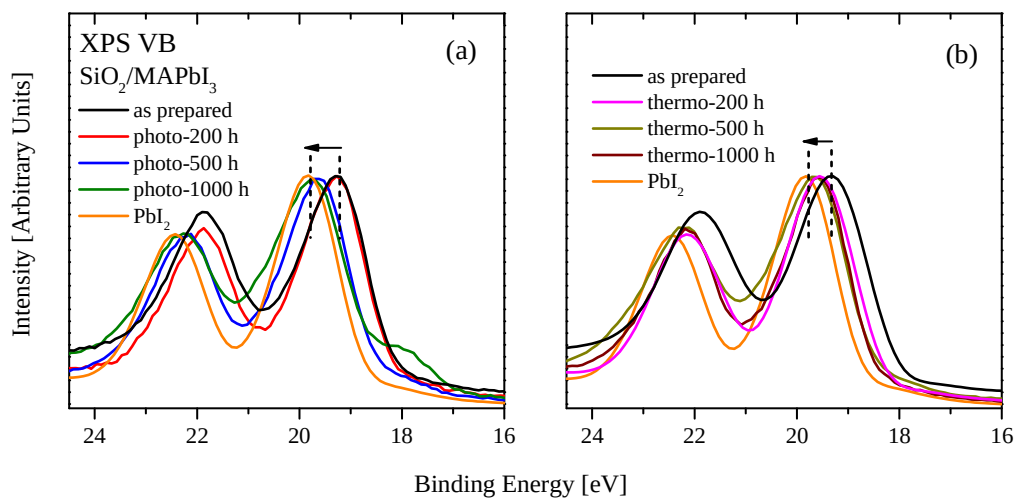

Fig. 3. XPS Pb 5d spectra of as-prepared, light irradiated (a) and annealed (b) SiO$_2$/MAPbI$_3$.

The next confirmation of the established patterns of XPS VB spectra behavior depending on the exposure time during photo irradiation and annealing are the data on the energy position of the XPS spectra of the Pb 5d-core levels shown in the high-energy part of Figure 2 (in the energy range of 17-24.5 eV) and on a larger scale in Fig. 3. As follows from these spectra, the binding energy of the XPS Pb 5d-core levels practically coincides for light irradiated at 200 hours and the initial sample, while a further increase in exposure time up to 500 hours leads to a high-energy shift, and continues at a dose of 1000 hours until full coincidence with the binding energy of the reference PbI$_2$ compound (Fig. 3a). The similar trends are observed for the annealed samples (Fig. 3b) only the threshold values of the onset of degradation already appear at a dose of 200 hours.

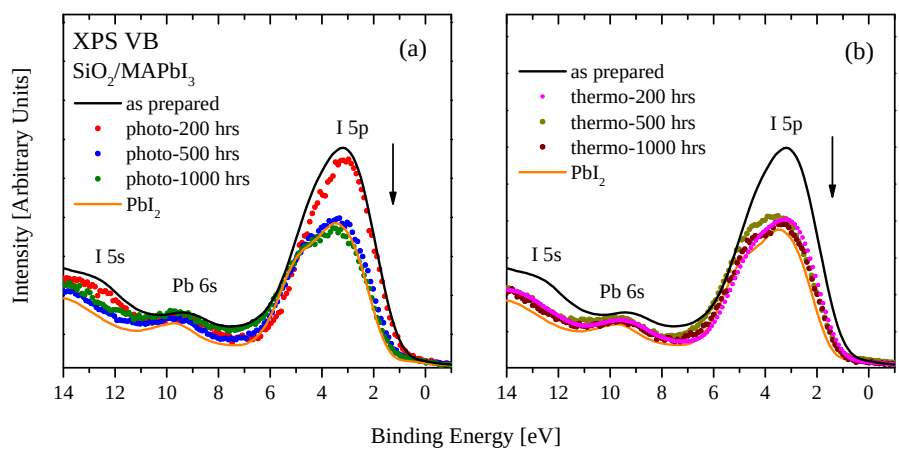

Fig. 4. XPS VB spectra of as-prepared, light irradiated (a) and annealed (b) SiO$_2$/MAPbI$_3$.

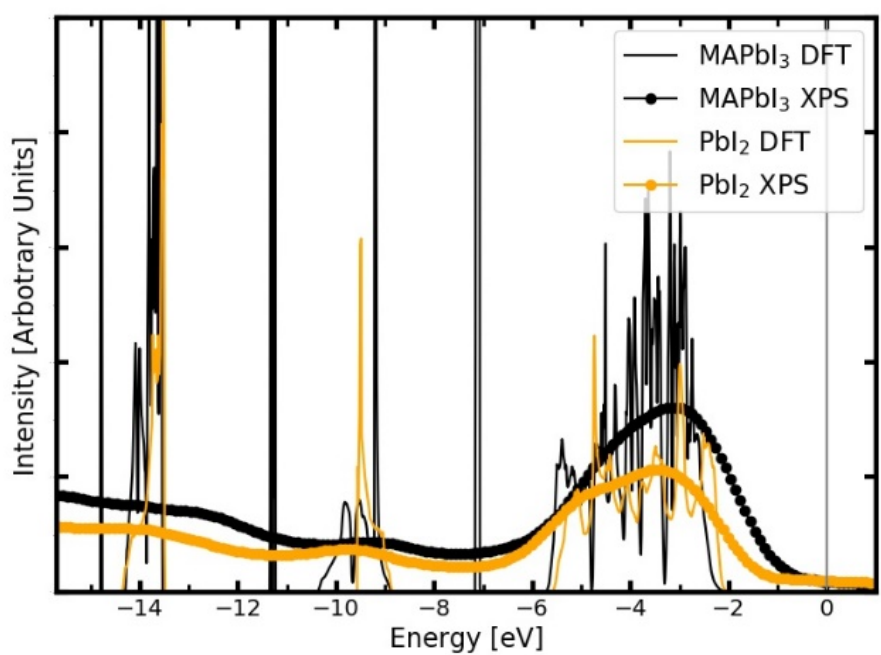

Fig. 5 Comparison of XPS and DFT valence band spectra for MAPbI$_3$ (black) and PbI$_2$ (orange). The experimental and theoretical data are shown by dots and lines, respectively.

Analysis of the elemental and orbital compositions of the molecular levels of the [CH$_3$NH$_3$]$^-$ organic cation showed that in the sequence of its levels shown in Fig. 5: -7, -9.1 - 11.3, -14.7 eV, the first displays the energy position of the CH$_3$ group (main contributions from H 1s and C 2p), the second - the binding energy of N-C bonds (N 2p and C 2p), and the third group - the binding energy of NH$_3$-group (N 2p and H 1s). Further at the larger negative energies (-14.7 eV) the levels with participation of C 2s and N 2s are located.. The main valence band (from -2.2 to -5.5 eV) is formed mainly by I 5p electrons. It is located closest to the $CH_3$ group (-7 eV) and has a small overlap with it due to the electrostatic interaction of a positively charged electric dipole with a negative charged $PbI_6$ octahedra.

The comparison of the experimental XPS valence band spectra of $MAPbI_3$ and $PbI_2$ and theoretical DFT calculations of these compounds presented in Fig. 5. shows overall good agreement. The XPS valence bands at -5.5÷-2.2 eV match well with its theoretical counterparts displaying a proper reduction of the I 5p intensities going from $MAPbI_3$ to $PbI_2$. I 5s narrow bands for both compounds are located about -14 eV and demonstrate the same change in relative intensities. Pb 6s bands lie at -9 eV and coincide with the experimental data. As stated before the methylammonium peaks are located at -14.7, -11.3, -9.1 and -7 eV. Due to weak hybridization of the $CH_3NH_3$ cation with the $PbI_6$ octahedron these peaks are very narrow, less than 0.2 eV. One should remind here about the experimental energy resolution of 0.5 eV, that leads to the absence of the methylammonium peaks in the experimental picture.

Summing up the research it should be noted that our measurements of XPS survey, 5d and valence band spectra showed that irradiation with visible light and heating (annealing) of $MAPbI_3$ perovskite lead to a systematic decrease in the I:Pb ratio, high-energy shift of XPS Pb 5d-lines and decrease of the contribution of I 5p-states. In all the cases listed above, the spectral characteristics obtained for the limiting aging time (1000 hours) are close to the $PbI_2$ compound that is the supposed product of decomposition upon irradiation and annealing. The decrease in the contribution of I 5p-states observed in the experiment up to coincidence with that in the $PbI_2$ compound is confirmed by DFT-calculations of the electronic structure, which also show a decrease in the density of I 5p-states during the transition from compound $MAPbI_3$ to $PbI_2$. Thus, the proposed mechanism of photo- and thermal degradation associated with the decomposition of the hybrid perovskite $MAPbI_3$ with $PbI_2$ phase separation was confirmed by the measurements of x-ray photoelectron spectra and DFT-calculations.

**Conclusion**

The main experimental result of this work is the establishment of the fact that the heating of the hybrid perovskite is a stronger degrading factor than exposure to the visible light. We assume. that this is due to the physical nature of the organic cation $[CH_3NH_3]^-$ which is an electrical dipole participating in an electrostatic dipole-dipole interaction with neighbor dipoles and a negatively charged framework of octahedra with $I^-$ ions. According to [14], the energies of these interactions are close to the energy of thermal phonons, therefore the probability of their absorption by the system can resonantly exceed the absorption of visible light, whose frequency is much higher than the frequency of phonons.

**Acknowledgements**

This work is supported by Russian Science Foundation (project 19-73-30020). The DFT calculations were supported by Ministry of Education and Science of Russia (task 3.7270.2017 / 8.9) and Theme "Electron" № AAAA-A18-118020190098-5.